\documentstyle[sprocl]{article}
\input{psfig.sty}
\def\Journal#1#2#3#4{{#1} {\bf #2}, #3 (#4)}

\def\NCA{\em Nuovo Cimento}

\def\PRD{{\em Phys. Rev.} D}

\def\be{\begin{equation}}
\def\ee{\end{equation}}
\def\bea{\begin{eqnarray}}
\def\eea{\end{eqnarray}}

\begin{document}
\tolerance=2000
\title{Theory of Relativistic Reference Frames for High-Precision Astrometric Space Missions }
\author{ Sergei M. Kopeikin\footnote{E-mail: kopeikins@missouri.edu} }
\address{Department of Physics \& Astronomy, University of Missouri-Columbia,\\
Columbia, Missouri 65211, USA}

\maketitle \abstracts{Recent modern space missions
deliver invaluable information about origin of our universe,
physical processes in the vicinity of black holes and other exotic
astrophysical objects, stellar dynamics of our galaxy, etc. On the
other hand, space astrometric missions make it possible to determine with
unparalleled precision distances to stars and cosmological objects
as well as their physical characteristics and positions on the
celestial sphere. Permanently growing accuracy of space
astronomical observations and the urgent need for adequate data
processing algorithms require corresponding development of an
adequate theory of reference frames along with unambiguous
description of propagation of light rays from a source of light to
observer. Such a theory must be based on the Einstein's general
relativity and account for numerous relativistic effects both in
the solar system and outside of its boundary. The main features of the relativistic theory of reference frames are presented in this work.
A hierarchy of the frames is described starting from the perturbed
cosmological Friedmann-Robertson-Walker metric and going to the
observer's frame through the intermediate barycentric and
geocentric frames in the solar system. Microarcsecond astrometry
and effects of propagation of light rays in time-dependent
gravitational fields are discussed as well. }

\section{Introduction}

The role of high quality reference frames in astronomy has been recognized early by both theorists and observers. Astrometric and navigation data rely on observations referred to a frame, either local or global. The choice of the reference frame may be driven by instrumental considerations or be based upon deeper theoretical grounds. This paper deals with the latter subject.

The original approaches to construct reference frames in astronomy
were completely based on the concepts of Newtonian gravity and
Euclidean absolute space and time \cite{km}. Modern astrometry,
however, is operating at the angular resolution already exceeding
1 milliarcsecond (see, for example, \cite{min}). At this level the
primary gravitational theory must be the General 
Relativity Theory (GRT) with a corresponding replacement of the Euclidean
space and time by the four-dimensional Riemannian space-time
manifold. In other words, the theoretical basis of modern astrometry
must be entirely relativistic. Recognition of this fact is rapidly
spreading in the astrometric community especially after the successful
completion of the HIPPARCOS mission \cite{hip}, materialization of the
International Coordinate Reference Frame in the sky, development of
technologically new ideas in space astrometry, and the adoption
by the XXIVth General Assembly (GA) of the International Astronomical
Union (IAU) of a new relativistic framework for the reference frames
in the Solar system \cite{iau}.

In what follows we shall describe in more detail the initial theoretical 
motivations that stimulated the development of modern relativistic framework for the reference frames in the solar system and new ideas driving the development 
of the theory beyond 2000. Then, in section 2, we give a description of 
the IAU-1991 reference frame framework and compare it in section 3 with the 
present day formulation adopted by the XXIVth GA of the IAU-2000. A possible way of matching the IAU-2000 framework with cosmological model of the universe 
is considered in section 4. The basic theoretical ideas for approaching microarcsecond 
accuracy in relativistic space astrometry is outlined in section 5. 

\subsection{Initial Motivations}

Transition from the Newtonian concepts towards more profound
theoretical relativistic approach for the construction reference
frames in modern astrometry and celestial mechanics can be traced
back to the period 1975-1992 and is associated with the breakthrough
in the solution of two major problems: development of a
self-consistent framework for derivation of the post-Newtonian
equations of translational and rotational motion of
self-gravitating extended bodies in the solar system
\cite{k88}$^,$ \cite{bk89} and development of the higher-order
relativistic celestial mechanics of binary pulsars
\cite{d}$^,$ \cite{s}$^,$ \cite{k85} discovered by Hulse and
Taylor \cite{ht}. The main concern of theorists working on relativistic
celestial mechanics of the solar system bodies was related to the
problem of unambiguous interpretation of astronomical measurements
and separation of small coordinate perturbations from the real
physical relativistic effects. Later on the question regarding the
best choice for the gauge conditions imposed on the metric tensor
arose, for a number of gauges were used in calculations by
various groups and there were arguments about their advantages and
disadvantages. But, probably the most serious was the problem of
construction of the local geocentric (and planetocentric)
reference frame in the post-Newtonian approximation(s). Apparently
there existed principal difficulties in solving these problems.
This was because the simplest approximation of the Earth's
gravitational field by the Schwarzschild solution could not be
considered as accurate enough due to the noticeable contribution
to the metric from the Earth's rotation and oblateness as well as the
existence of tidal forces from the Moon, Sun, and other planets. It
was also recognized \cite{k88} that the well-known procedure of
construction of the Fermi normal coordinates \cite{ber} can not be
applied due to the ambiguity in choosing the background space-time
manifold and deviation of the Earth's center-of-mass world line
from geodesic motion. Lack of a rational approach to the problem
led to the fact that a simple (Euclidean-like) spatial translation of the
origin of the barycentric reference frame to the Earth's geocenter
was used in order to construct a geocentric reference frame at the
post-Newtonain approximation. However, though such a procedure is
allowed in GRT because of the coordinate freedom, the Euclidean
translation to the geocenter produces large relativistic effects
having pure coordinate origin and, hence, unobservable. In
addition, in such a frame the geometric shape of the Earth moving
along the elliptic orbit undergoes the Lorentz and gravitational
contractions which must be compensated by spurious internal
stresses in order to prevent the appearance of unphysical
deformations. Similar problems were also met in developing general
relativistic celestial mechanics of binary pulsars \cite{d300}.

\subsection{Motivations in the New Millennium}

Recent technological developments make it necessary to extend  
the domain of applicability of the relativistic theory of 
reference frames outside the boundaries of the solar system. 
New generation of astrometric satellites which include FAME \cite{fame}, SIM \cite{sim},
and GAIA (recently adopted as a cornerstone mission of ESA
\cite{p}) requires an absolutely new approach for unambiguous
interpretation of astrometric data obtained from the on-board
optical interferometers. FAME's resolution is about 100
microarcseconds for stars having 10-th stellar
magnitude. The resolution of GAIA for the same kind of stars is
expected to be already 100 times better. At this level of accuracy
the problem of propagation of light rays must be based on the
recently developed post-Minkowskian "Lorentz-covariant" approximation \cite{ks} that allows us to
integrate the equations of light propagation without artificial
assumptions about the motion of light-ray-deflecting bodies.
Besides, the treatment of the parallax, aberration, and proper motion
of celestial objects becomes much more involved requiring better
theoretical definitions of reference frames on a curved space-time
manifold.

Additional motivation for improving  relativistic  theory of reference 
frames is related to the problem of calculation of the incoming signals of the 
space interferometric gravitational wave detectors 
like LISA \cite{lisa} which will consist of three satellites flying in space separated 
by distances of order $~1.5\times 10^6$ km. Detection of gravitational waves can be 
done only under the condition that all coordinate-dependent phenomena are completely 
understood and subtracted from the signal. On the other hand, LISA will fly in 
the near-zone of the Sun which is considered as a source of low-frequency gravitational 
waves produced by oscillations of its interior (known as g-modes) \cite{curt}. Motion of the satellites
carrying lasers and mirrors as well as
propagation of light rays along the baseline of LISA in the field of such g-modes must be carefully 
studied in order to provide a correct interpretation of observations.

Many other relativistic experiments for testing GRT and alternative 
theories of gravity also demand advanced theory of relativistic reference 
frames (see, for example, \cite{ciuf}$^,$ \cite{gpb}$^,$ \cite{llr}). 

\subsection{Historical Remarks}

Historically it was de Sitter \cite{desit1}$^,$ \cite{desit2} who worked out a relativistic approach to build reference 
frames in GRT. He succeeded in the derivation of relativistic equations of motion of the solar system bodies 
and discovered the main post-Newtonian effects including gravitomagnetic perihelion precession 
of a planetary orbit due to the angular 
momentum of the Sun as well as the geodetic precession ("de Sitter-Fokker effect") that 
has been verified with the precision of about $1\%$ \cite{berc}$^,$ \cite{sh}. These effects are 
essential in the present-day definition of dynamical or kinematical rotation of a reference frame \cite{bk89}. Later on 
Lense and Thirring gave a more general treatment of the gravitomagnetic dragging of a satellite 
orbiting around a massive rotating body \cite{lt} but hardly believed that the effect can be 
measured in practice. 

It was Ginzburg \cite{g}  who first realized that the Lense-Thirring 
effect can indeed be measured using artificial satellites and considered 
the problem of separation of coordinate and physical effects. But only recently the 
experimental verification of the 
Lense-Thirring effect has come about \cite{ciuf}. 

Ginzburg's paper and the success of the soviet space program motivated Brumberg \cite{b} to develop 
a post-Newtonian Hill-Brown theory of 
motion of the Moon in the solar barycentric coordinate system. Baierlein \cite{ba} extended  
Brumberg's approach accounting for the eccentricity of Earth's orbit. 

A different original approach to 
the problem of motion of the Moon and the construction of reference frames in GRT was suggested by 
Mashhoon and Theiss in a series of papers (see, for example, \cite{mt} and references 
therein). Instead of making use of the post-Newtonian approximations (PNA) they developed a 
post-Schwarzschild treatment of gravitomagnetic effects in the three-body problem. It allowed 
them to discover that the validity of PNA is restricted in time and the geodetic and gravitomagnetic precessions 
are parts of the more general phenomena involving a long-term relativistic nutation 
("Mashhoon-Theiss effect"). The same authors also introduced a precise definition of the 
local geocentric frame with the Earth considered as a massive monopole particle \cite{mt}$^,$ \cite{mash}. 

A more general approach to the problem of construction of reference frames in GRT was initiated 
in \cite{k88} (see also \cite{k87}), where the matched asymptotic expansion technique and a decomposition of gravitational fields of bodies in the (Newtonian) multipoles were employed for 
this purpose.
Our approach has been further used for the development of the extended Brumberg-Kopeikin (BK) 
formalism for building relativistic astronomical frames. This formalism relies upon 
Einstein's equations, which are solved for the construction of local geocentric and global 
barycentric reference systems, and matching technique, which is used for setting up 
relativistic transformations between these two systems \cite{bk89}$^,$ \cite{bk89a}. 

T. Damour, M. Soffel and C. Xu (DSX) have extended the Brumberg-Kopeikin 
theory by applying the post-Newtonian definitions of the ("Blanchet-Damour") gravitational 
multipole moments \cite{dsx}. Elements of the BK formalism were introduced in resolutions of the GA of the IAU in 1991 \cite{a4}. The complete BK-DSX theory is presently accepted by the XXIVth 
GA of the IAU-2000 as a basic framework for setting up relativistic time scales and 
astronomical reference frames in the solar system.

\section{The IAU-1991 Reference Systems Framework}

Official transition of the astronomical community from Newtonian
positions to relativistic concepts began in 1991 when a few 
recommendations (resolution A4) were adopted by the GA of the IAU.

In the first recommendation, the metric tensor in space-time
coordinates $(ct,{\bf x})$ centered at the barycenter of an
ensemble of masses is recommended to be written in the form
\begin{eqnarray}
g_{00} &=& -1
+ {2U(t,{\bf x}) \over c^{2}} + O(c^{-4})\;,\label{1}\\
\label{2}
g_{0i} &=& O(c^{-3})\;,\\
\label{3}
g_{ij} &=& \delta_{ij}\,\left[1+{2U(t,{\bf x}) \over c^{2}}\right] +O(c^{-4})\;,
\end{eqnarray}\noindent
where $c$ is the speed of light in vacuum,
$U$ is the sum of the gravitational potentials of
the ensemble of masses, and of a tidal potential generated by bodies
external to the ensemble, the latter  vanishing at the
barycenter. This recommendation recognizes that space-time cannot be described by
a single coordinate system. The recommended form of the metric tensor
can be used not only to describe the barycentric celestial
reference system (BCRS) of the whole
solar system, but also to define the geocentric
celestial reference system (GCRS)
centered on the center of mass of the Earth
with a suitable function $\hat{U}$, now depending upon geocentric coordinates.
In
analogy to the GCRS a corresponding celestial
reference system can be constructed
for any other body of the Solar system.

In the second recommendation, the origin and orientation of the
spatial coordinate grids for the solar system (BCRS) and for the Earth
(GCRS) are defined.
The third recommendation defines $TCB$ (Barycentric Coordinate Time) and
$TCG$ (Geocentric Coordinate Time) -- the time coordinates of the BCRS and
GCRS, respectively. The relationship
between $TCB$ and $TCG$ is given by a full 4-dimensional transformation\noindent
\begin{equation}\label{4}
TCB - TCG= c^{-2}\,
\left[\int^{t}_{t_0}
\left({v_{E}^{2} \over 2}
      + \overline{U}(t,{\bf x}_E) \right)  dt
+ v^i_E r^i_E \right] + O(c^{-4}),
\end{equation}\noindent
where $(x^i_{E})\equiv {\bf x}_E(t)$ and $(v^i_{E})$ are the barycentric coordinate position
and velocity of the geocenter, $r^i_E = x^i - x^i_{E}$ with ${\bf x}$ being the
barycentric position of the observer, and $\overline{U}(t,{\bf x}_E)$
is the Newtonian potential of all solar system bodies evaluated at the geocenter apart from that of the Earth.

In August 2000, a new resolution B1 on
reference frames and time scales in the solar system was
adopted by the XXIVth General Assembly of the IAU \cite{iau}. The resolution is based on the first post-Newtonian
approximation of GRT and completely abandons the Newtonian point of
view on space and time \footnote{ How the resoltuion B1 was prepared can be traced in the materials of the corresponding working groups of the IAU \cite{rcma}$^,$ \cite{jcr}}.  

\section{The IAU-2000 Reference Systems Framework}
\subsection{Conventions for the Barycentric Celestial Reference System}
BCRS is defined mathematically in terms of a metric tensor which reads
\begin{eqnarray}
g_{00} &=& - 1 + {2 w \over c^2} - {2w^2 \over c^4} + O(c^{-5})\;,
\label{5} \\
g_{0i} &=& - {4 \over c^3} w_i + O(c^{-5})\;,\label{6}
 \\
g_{ij} &=& \delta_{ij}
\left( 1 + {2 \over c^2} w \right) + O(c^{-4})\; .
\label{7}
\end{eqnarray}\noindent
Here, the post-Newtonian gravitational potential $w$ generalizes the
usual Newtonian potential $U$ and $w^i$ is the vector potential related
with gravitomagnetic effects.

This form of the barycentric metric tensor implies that the barycentric
spatial coordinates $x^i$ satisfy the harmonic gauge condition.
The main arguments in favor of the harmonic gauge are: (1)
tremendous work on GRT has been done with the harmonic
gauge that was found to be a useful and simplifying gauge for all kinds
of applications, and (2)
in contrast to the standard post-Newtonian (PN) gauge (see, for example, \cite{will}) the harmonic gauge can be defined
to higher PN-orders, and in fact for the exact Einstein theory of
gravity.

Assuming space-time to be asymptotically flat (no gravitational fields exist
at infinity) in the standard harmonic gauge the post-Newtonian
field equations of GRT are solved by
\begin{eqnarray}
w(t,{\bf x}) &=& G \int d^3 x' \;
{\sigma(t, {\bf x}') \over \vert {\bf x} - {\bf x}' \vert}
 + {G \over 2c^2}  {\partial^2 \over \partial t^2}
\int d^3 x' \, \sigma(t,{\bf x}') \vert {\bf x} - {\bf x}' \vert \; ,
\label{8}\\\label{9}
w^i(t,{\bf x}) &=& G \int d^3 x'\; {\sigma^i (t,{\bf x}') \over
\vert{\bf x} - {\bf x}' \vert }\;,
\end{eqnarray}\noindent
where $
\sigma(t,{\bf x}) = c^{-2}(T^{00} + T^{ss}),$ $
\sigma^i(t,{\bf x}) = c^{-1}T^{0i}$, and
$T^{\mu\nu} = T^{\mu\nu}(t,x^i)$ are the components of the stress-energy tensor in the barycentric coordinate system, $T^{ss}
= T^{11} + T^{22} + T^{33}$.

\subsection{Conventions for the Geocentric Celestial Reference System}

GCRS is defined in terms of the geocentric metric tensor
\begin{eqnarray}
G_{00} &=& - 1 + {2 W \over c^2} - {2W^2 \over c^4} + O(c^{-5}),
\label{11} \\\label{12}
G_{0a} &=& - {4 \over c^3} W_a,
 \\
G_{ab} &=& \delta_{ab}
\left( 1 + {2 \over c^2} W \right) + O(c^{-4}) \, .
\label{13}
\end{eqnarray} \noindent
Here $W = W(T,{\bf X})$ is the post-Newtonian gravitational potential in
the geocentric system and $W^a(T,{\bf X})$ is the corresponding vector
potential.
These geocentric potentials should be split into two parts: potentials
$W_E$ and $W_E^a$ arising from the gravitational action of the Earth
and external parts $W_{\rm ext}$ and $W^a_{\rm ext}$ due to tidal and
kinematic effects. The external parts are assumed to vanish at the
geocenter
and admit an expansion into positive powers of ${\bf X}$.
Explicitly,
\begin{eqnarray}
W(T,{\bf X}) &=& W_E(T,{\bf X})
+ W_{\rm kin}(T,{\bf X})+W_{\rm tidal}(T,{\bf X})
\;,
\label{14}\\
\label{15}
W^a(T,{\bf X}) &=& W^a_E(T,{\bf X})
+ W^a_{\rm kin}(T,{\bf X})+W^a_{\rm tidal}(T,{\bf X})
\;.
\end{eqnarray}\noindent
The Earth's potentials  $W_E$ and $W^a_E$
are defined in the same way as $w_E$ and $w_E^i$ but with quantities
calculated in the GCRS. $W_{\rm kin}$ and $W_{\rm kin}^a$ are kinematic contributions that are linear in $X^a$
\begin{equation}\label{16}
W_{\rm kin}= Q_b X^b,
\qquad
W^a_{\rm kin} = {1\over 4}\;c^2
\varepsilon_{abc} (\Omega^b - \Omega^b_{\rm prec})\;X^c\; ,
\end{equation}\noindent where
$Q_b$ characterizes the deviation of the actual worldline of
the origin of the GCRS from geodesic motion in the
external gravitational field (for more details see \cite{k88}$^,$ \cite{bk89}$^, $ \cite{dsx})
\begin{equation}\label{17}
Q_b=\partial_b w^{\rm ext}({\bf x}_E)-a_E^b+O(c^{-2})\;.
\end{equation}\noindent
Here
$a_E^i={dv^i_E/dt}$ is the
barycentric acceleration of the
origin of the GCRS (geocenter). The function
$\Omega^a_{\rm prec}$ describes the relativistic precession
of dynamically nonrotating spatial axes with respect to remote objects:
\begin{eqnarray}\label{18}
\Omega_{\rm prec}^i =
{1\over c^2}\,
\varepsilon_{ijk}\,
\left(
-{3\over 2}\,v^j_E\,\partial_k w^{\rm ext}({\bf x}_E)
+2\,\partial_k w^{j}_{\rm ext}({\bf x}_E)
-{1\over 2}\,v^j_E\,Q^k
\right).
\end{eqnarray}\noindent
The three terms on the right-hand side of this equation
represent the geodetic, Lense-Thirring, and Thomas precessions,
respectively.  One can prove that $\Omega^a_{\rm iner}$ is dominated by
geodetic precession  amounting to $\sim 2^{\prime\prime}$ per century
plus short-periodic terms usually called geodetic nutation.
One sees that for $\Omega^a = \Omega^a_{\rm prec}$ the vector potential
$W^a_{\rm kin}$ vanishes. This implies that dynamical equations of
motion of a test body, e.g., a satellite orbiting around the Earth, do not contain the
Coriolis and centrifugal terms, i.e., the local geocentric spatial
coordinates $X^a$ are {\it dynamically non-rotating}. For practical reasons,
however, the use of {\it kinematically non-rotating} geocentric
coordinates defined by $\Omega^a = 0$ is recommended.

Potentials $W^{\rm tidal}$ and $W_a^{\rm tidal}$ are generalizations
of the Newtonian tidal potential. 
We also note that the local gravitational potentials $W_E$ and $W_E^a$ of the
Earth are related to the barycentric gravitational potentials $w_E$ and
$w^i_E$ by the relativistic transformations \cite{iau}.

\subsection{Transformations between the reference systems}
The coordinate transformations between the BCRS and GCRS are written as
\begin{eqnarray}
\label{22}
T&=&t - {1\over c^2} \left[ A(t) + v_E^i\,r_E^i \right]
\\\nonumber
&&\ \ \ \ \
+ {1\over c^4} \left[ B(t) + B^i(t)\,r_E^i +
B^{ij}(t)\,r_E^i\,r_E^j + C(t,{\bf x}) \right] +O(c^{-5}),
\\\label{23}
X^i&=&
r^i_E+\frac 1{c^2}
\left[\frac 12 v_E^i v_E^j r_E^j + w_{\rm ext}({\bf x}_E) r^i_E
+ r_E^i a_E^j r_E^j-\frac 12 a_E^i r_E^2
\right]+O(c^{-4}),
\end{eqnarray}\noindent
where functions $A(t), B(t), B^i(t), B^{ij}(t), C(t,{\bf x})$ can be found in \cite{iau}.
Let us also remark that the harmonic gauge
condition does not fix the function $C(t,{\bf x})$ uniquely. However, we
prefer to fix it in the time transformation for practical reasons.

\section{Matching the IAU-2000 Framework with Cosmological Reference Frame}

The rapidly growing accuracy of astronomical measurements makes it necessary to take into account some important cosmological effects for an adequate interpretation of optical and radio observations of cosmological lenses, anisotropy in cosmic microwave background radiation, etc. For this reason the matching of the IAU-2000 framework for reference systems in the solar system with the cosmological reference frame becomes vitally important. In this section we outline the main ideas in this matching for the \underline{spatially flat} Friedmann-Robertson-Walker (FRW) cosmological model.

First, we consider the perturbation $h_{\alpha\beta}$ of the gravitational field as
\begin{equation}
\label{c1}
g_{\alpha\beta}=a^2(\tau)(\eta_{\alpha\beta}+h_{\alpha\beta})\;,\qquad\quad
\gamma^{\alpha\beta}=-h^{\alpha\beta}+{1\over 2}\eta^{\alpha\beta}h\;,\qquad h=\eta^{\alpha\beta}h_{\alpha\beta}\;,
\end{equation}
and impose the quasi-harmonic cosmological gauge conditions \footnote{These conditions were also independently discovered in the paper \cite{vega} for the case of de Sitter space-time.}  
\begin{equation}
\label{c2}
\partial_{\beta} \gamma^{\alpha\beta}=-2H\left(\gamma^{\alpha 0}-{1\over2}\eta^{\alpha 0}\gamma\right)\;,\qquad\quad
H(\tau)={\dot{a}\over a}\;,
\end{equation}
which eliminate almost all first order derivatives of the metric perturbation. 
Then, the linearized Einstein equations read
\begin{eqnarray}\label{c3}
\Box\gamma^{\alpha\beta}-2H\;\partial_\tau\gamma^{\alpha\beta}+2(2\dot{H}+H^2)h^{\alpha\beta}&+&4(\dot{H}-2H^2)\eta^{0(\alpha}h^{\beta)0}
\\\nonumber
&-&2(\dot{H}-H^2)\eta^{\alpha\beta}h^{00}=16\pi a^4\delta T^{\alpha\beta}\;,
\end{eqnarray}
where a dot over "the Hubble parameter" $H$ denotes the time derivative and $\delta T^{\alpha\beta}$ is the tensor of energy-momentum of perturbing source ("solar system").

For the particular case of matter-dominated background FRW cosmological model 
the reduced linearized Einstein equations (\ref{c3}) read
\begin{eqnarray}\label{c5}
\Box_g w-8H^2w&=&-4\pi a^4(\tau)\;\delta( T^{00}+ T^{kk})\;,
\\\label{c6}
\Box_g(\gamma+{1\over2}w)&=&-4\pi a^4(\tau)\;\delta( T^{00}-{7\over 2} T)\;,\\
\label{c7}
\Box_g w^{i}-5H^2 w^{i}&=&-4\pi a^4(\tau)\;\delta T^{0i}\;,\\
\label{c8}
\Box_g w^{ij}&=&-4\pi a^4(\tau)\;\delta T^{<ij>}\;,
\end{eqnarray}
where $w=-\left(\gamma^{00}+\gamma^{kk}\right)/4\;, w^i=-\gamma^{0i}/4$, $w^{ij}=-\gamma^{<ij>}/4$ \footnote{The square brackets around spatial indices denote symmetric and trace-free tensor \cite{kip}}, and $H(\tau)=2/\tau$. For the differential operator $\Box_g$ one has
\begin{equation}
\label{c9}   \Box_g\equiv\Delta-{\partial^2\over\partial\tau^2}-{4\over\tau}{\partial\over\partial\tau}\;,
\end{equation}
so that the solution of the inhomogeneous equation
$\Box_g F(\tau,{\bf x})=-4\pi a^4(\tau)\delta T(\tau,{\bf x}),$
can be found making use of the replacement \cite{sachs}
\begin{equation}
\label{c11}
F(\tau,{\bf x})={1\over\tau}{\partial\over\partial\tau}\left({ \Psi(\tau,{\bf x})\over\tau}\right)\;.
\end{equation}
transforming the equation in question into 
\begin{equation}
\label{c12}
{1\over\tau}{\partial\over\partial\tau}\left({\Box_\eta\Psi(\tau,{\bf x}) \over\tau}\right)
=-4\pi a^4(\tau) \delta T(\tau,{\bf x})\;,\qquad \Box_\eta=\Delta-\partial^2_\tau\;,
\end{equation}
which has a simple particular solution
\begin{equation}\label{c13}
\Psi(\tau,{\bf x})=\tau\int{d^3{\bf x}'\over |{\bf x}-{\bf x}'|}\int^{\tau'}_{1} s a^4(s) \delta T(s,{\bf x}')ds 
-\int d^3{\bf x}'\int^{\tau'}_{1} s a^4(s)\delta T(s,{\bf  x}')ds \;,
\end{equation}
where $\tau'=\tau-|{\bf x}-{\bf x}'|$ is a retarded time in a flat space-time.

Matching this solution with that defined in the BCRS of the IAU-2000 framework as flat-space retarded potentials is achieved after the replacement $\tau=1+H_R t$ and expanding all quantities depending on $\tau$ in the neighbourhood of the present epoch $\tau=1$ along with making use of the tensor transformation law for metric tensor. Results of the matching will be described elsewhere. 
\section{Relativistic Microarcsecond Astrometry}
The IAU-2000 framework for reference frames requires new advanced theory of astrometric data analysis. The key issue in this theory is solution of the problem of propagation of light rays with account for as many relativistic effects as required for making unambigious interpretation of the data.
Previous approaches for calculating light ray propagation in the framework of relativistic astrometry were based on making use either
metric tensor of exact solutions of GRT (Schwarzschild, Kerr, etc.), or
metric tensor of the post-Newtonian approximation (PNA), or
metric tensor of the plane weak quadrupolar gravitational wave.
All these approaches have difficulties and/or inconsistencies in describing light propagation at the microarcsecond threshold. 
The question arises how to avoid these difficulties and what is the proper formalism to deal with the problem. Treatment of this problem has been given recently in the framework of the "Lorentz-covariant" theory of light propagation \cite{ksge}$^-$\cite{kma}. 
and the formalism developed has an unrestricted ability to make calculations at any desired order of approximation with respect to the small parameter $v_a/c$ as well as many other advantages making it a powerful tool for theoretical predicitons of various relativistic effects being detectable by modern astrometric technique (see, for example, \cite{kg}$^-$ \cite{sazh}).

\section{Acknowledgments}
I appreciate the help of Bahram Mashhoon (UMC) in giving valuable references and for taking part in numerous discussions. Interesting conversations with Juan Ramirez (Universidad Complutense de Madrid) and Mikhail Sazhin (Sternberg Astronomical Institute) clarified some important issues for establishing the proper framework for the cosmological reference frames.

\end{document}